\documentclass[letterpaper, 10 pt, conference]{ieeeconf}  

\IEEEoverridecommandlockouts                              
\usepackage{graphics} 
\usepackage{epsfig} 
\usepackage{times} 
\usepackage{amsmath} 
\usepackage{amssymb}  
\usepackage{graphicx}      
\usepackage{amsfonts,mathtools} 
\usepackage{xcolor,epstopdf}
\usepackage{booktabs}
\usepackage[T1]{fontenc}
\usepackage[latin9]{inputenc}
\usepackage[caption=false,font=footnotesize]{subfig}    
\usepackage{xcolor}
\usepackage{epstopdf}
\usepackage{caption}
\usepackage{multirow}
\usepackage{soul}
\sethlcolor{green}

\newtheorem{theorem}{Theorem}

\newtheorem{proposition}{Proposition}
\newtheorem{remark}{Remark}

\newtheorem{assumption}{Assumption}

\title{\LARGE \bf Geometric Nonlinear Filtering with Almost Global Convergence for Attitude and Bias Estimation on the Special Orthogonal Group}

\author{Farooq Aslam, M. Farooq Haydar, and Suhail Akhtar
\thanks{Farooq Aslam, Muhammad Farooq Haydar, and Suhail Akhtar are with the Institute of Space Technology, Islamabad, Pakistan (E-mail address: {\tt\small farooq.aslam87@gmail.com}; {\tt\small muhammadfarooq.haydar@polimi.it}; {\tt\small suhail.akhtar} {\tt\small @ist.edu.pk})}
}

\begin{document}

\maketitle
\thispagestyle{empty}
\pagestyle{empty}

\begin{abstract}
This paper proposes a novel geometric nonlinear filter for attitude and bias estimation on the Special Orthogonal Group $SO(3)$ using matrix measurements.
The structure of the proposed filter is similar to that of the continuous-time deterministic multiplicative extended Kalman filter (MEKF).
The main difference with the MEKF is the inclusion of curvature correction terms in both the filter gain and gain update equations.
These terms ensure that the proposed filter, named the Generalized $SO(3)$-MEKF, renders the desired equilibrium of the estimation error system to be almost globally uniformly asymptotically stable (AGUAS). More precisely, the attitude and bias estimation errors converge uniformly asymptotically to zero for almost all initial conditions except those where the initial angular estimation error equals $\pi$ radians.
Moreover, in the case of small estimation errors, the proposed generalized $SO(3)$-MEKF simplifies to the standard $SO(3)$-MEKF with matrix measurements.
Simulation results indicate that the proposed filter has similar performance compared to the latter.
Thus, the main advantage of the proposed filter over the MEKF is the guarantee of (almost) global uniform asymptotic stability.
\end{abstract}



\section{Introduction}\label{sec:introduction}
{A}{ttitude} estimation methods capable of obtaining accurate estimates from noisy and biased sensor measurements are vital components in aerospace systems. Due to their practical and theoretical significance, attitude estimation algorithms have been studied extensively and a wide array of methods have been developed, ranging from algebraic attitude determination methods to fixed-gain observers, extended Kalman filters (EKF), and other nonlinear filtering algorithms \cite{crassidis2007survey}.

In the case of fixed-gain observers, a seminal result is the nonlinear complementary filter proposed in \cite{mahony2008nonlinear} which uses fixed gains, is easy to implement, and has steady-state performance comparable to that of the multiplicative extended Kalman Filter (MEKF). 
In contrast to fixed-gain observers, the MEKF applies the extended Kalman filtering (EKF) methodology to the estimation error system obtained using the multiplicative quaternion error \cite{lefferts1982kalman,markley2003attitude}.
An extension of the Kalman filter for linear systems, the MEKF remains the industry standard for attitude estimation applications. More generally, the EKF framework has also been used in combination with invariant estimation errors to obtain the Invariant Extended Kalman Filter (IEKF) for kinematic systems on matrix Lie groups \cite{bonnabel2007left,barrau2014intrinsic,barrau2016invariant}. In this more general framework, the standard (quaternion) MEKF can be interpreted as a left-invariant extended Kalman filter \cite{zamani2015nonlinear,gui2018quaternion}. 
Similarly, Mortensen's maximum likelihood recursive nonlinear filtering methodology \cite{mortensen1968maximum} has been used by Zamani et al. to obtain deterministic minimum energy filters for attitude estimation \cite{zamani2011near,zamani2012second,ZamaniThesis}.

An advantage of fixed-gain observers, such as the nonlinear complementary filter \cite{mahony2008nonlinear}, is that they possess (almost) global stability guarantees. 
On the other hand, filters with time-varying gains typically do not have such strong guarantees. 
A case in point is the MEKF for which almost global asymptotic stability (AGAS) has been established using unbiased velocity measurements \cite{jensen2011generalized}. 
When bias estimation is considered as well, the MEKF has, to the best of our knowledge, only local stability guarantees. 
In a similar vein, the IEKF is also known to be a locally asymptotic observer. 
In contrast, the combination of a fixed-gain observer and the MEKF results in a two-stage cascaded filter, known as the multiplicative exogenous Kalman filter (MXKF), which has AGAS guarantees \cite{stovner2018attitude}.
As this paper will demonstrate, an alternative path to global stability can be obtained by making a few crucial modifications to the filter gain and gain update equations of the MEKF, leading to a novel geometric nonlinear filter, with (almost) global stability guarantees, for attitude and bias estimation on $SO(3)$.

In particular, the filter proposed in this paper has two salient properties. 
Firstly, in the case of small estimation errors, the proposed filter reduces to the conventional $SO(3)$-MEKF (with matrix measurements). For this reason, the proposed filter is referred to as the Generalized $SO(3)$-MEKF.
Secondly, the proposed filter renders the desired equilibrium of the estimation error system to be almost globally uniformly asymptotically stable (AGUAS). More precisely, the attitude and bias estimation errors converge uniformly asymptotically to zero for almost all initial conditions except those where the initial attitude estimation error has an angle of rotation equal to $\pi$ radians.
The convergence guarantees are made possible by the inclusion of curvature correction terms in the equations for the MEKF.
The rest of the discussion is structured as follows. After Section \ref{sec:preliminaries} details various mathematical relations that will be used in the filter development, Section \ref{sec:ProblemFormulation} sets up the filtering problem to be addressed in the main body of the paper. Thereafter, Section \ref{sec:GeneralizedSO3MEKF} derives the Generalized $SO(3)$-MEKF and analyzes its stability. Lastly, Section \ref{sec:SimulationResults} presents simulation results demonstrating the performance of the proposed filter. 

\section{Preliminaries}\label{sec:preliminaries}


For $x,y\in\mathbb{R}^3$, the \emph{cross} $(\times)$ operator returns the skew-symmetric matrix $x^{\times}$ such that $x^{\times}y=x\times y$, that is,
\[
x^{\times}={\begin{bmatrix}x_1\\x_2\\x_3\end{bmatrix}}^{\times}\triangleq\begin{bmatrix}0&-x_3&x_2\\x_3&0&-x_1\\-x_2&x_1&0\end{bmatrix}.
\]
The inverse of the cross operator, denoted by the \emph{vee} $(\vee)$ map, recovers the entries of the vector $x$ from $x^{\times}$, that is, $(x^{\times})^{\vee}=x$. The Special Orthogonal Group $SO(3)$ refers to the group of rotation matrices with dimension $3$, that is,
\[
SO(3)\triangleq\{R\in\mathbb{R}^{3\times3}:RR^{\top}=R^{\top}R=I,\det(R)=1\}.
\]
The Rodrigues formula allows us to express $R\in SO(3)$ as
\begin{equation}
R(\alpha,\theta)=I+\left(\sin\theta\right)\alpha^{\times}+\left(1-\cos\theta\right)\alpha^{\times}\alpha^{\times},\label{eq:Rtilde_axang}
\end{equation}
where $(\alpha,\theta)\in S^2\times\mathbb{R}$ denote the axis of rotation and the angle of rotation associated with the rotation matrix $R$. The chordal (or Euclidean) metric on $SO(3)$ is defined as
\begin{equation}
\Psi_c(R) \triangleq \text{tr}[I-R]=2(1-\cos\theta).\label{eq:ChordalMetric}
\end{equation}
In addition, we consider two error vectors associated with the chordal metric. The first of these vectors is the left-trivialized derivative of $\Psi_c$ with respect to $R$, and is obtained as
\begin{equation}
\psi(R)\triangleq\frac{1}{2}(R-R^{\top})^{\vee}=(\sin\theta)\alpha.\label{eq:psi}
\end{equation}
The second error vector, associated with the chordal metric, is defined as
\begin{equation}
e_R(R)\triangleq\frac{1}{\sqrt{1+\text{tr}[R]}}(R-R^{\top})^{\vee}=\frac{2\sin\theta}{\sqrt{2+2\cos\theta}}\alpha.\label{eq:eR}
\end{equation}
An important property of $e_R$ is that its squared $2$-norm equals the chordal metric. In particular, using the axis-angle representation in \eqref{eq:ChordalMetric} and \eqref{eq:eR}, we observe that:
\begin{align}
\Vert e_{R}\Vert^{2}&=\frac{4\sin^{2}\theta}{2+2\cos\theta}=\frac{4(1+\cos\theta)(1-\cos\theta)}{2(1+\cos\theta)}\nonumber\\
&=2(1-\cos\theta)\nonumber\\
\implies\Vert e_{R}\Vert^{2}&=\text{tr}[I-R].
\end{align}
In addition, we consider two matrices related to the error vectors $\psi$ and $e_R$. These are defined as follows:
\begin{subequations}
\begin{align}
E_c(R) &\triangleq \frac{1}{2}\left(\text{tr}[R]I-R^{\top}\right), \quad R\in SO(3),\label{eq:Ectilde} \\
E(R) &\triangleq \frac{1}{\sqrt{1+\text{tr}[R]}}\left(2E_c(R)+\frac{1}{2}e_Re_R^{\top}\right).\label{eq:E}
\end{align}
\end{subequations}
From the analytical framework developed in \cite{lee2012exponential}, we observe that the eigenvalues of the product $E(R)E^{\top}(R)$ are $\{1,1,(1+\cos\theta)/2\}$. We obtain the determinant of $E(R)$ as
\begin{equation}
\det{(E(R))}=\sqrt{\frac{1+\cos\theta}{2}},
\end{equation}
and note that $E(R)$ is invertible for $\cos\theta\neq-1$, that is, for $\text{tr}[R]\neq-1$. In order to obtain the matrix inverse of $E(R)$, we use the axis-angle representation \eqref{eq:Rtilde_axang} to re-state $E(R)$ as
\begin{equation}
E(R)=\frac{1}{\sqrt{2+2\cos\theta}}[(1+\cos\theta)I+\left(\sin\theta\right)\alpha^{\times}].\label{eq:E_axang}
\end{equation}
For $\cos\theta\neq-1$, we obtain the matrix inverse as
\begin{align}
E^{-1}(R)&=\frac{1}{\sqrt{2+2\cos\theta}}[(1+\cos\theta)I-\left(\sin\theta\right)\alpha^{\times}+\cdots \nonumber \\
&\quad\quad+(1-\cos\theta)\alpha\alpha^{\top}].\label{eq:E_inv_axang}
\end{align}
The corresponding rotation matrix representation, for $\text{tr}[R]\neq-1$, is given by
\begin{align}
E^{-1}(R) &= \frac{1}{\sqrt{1+\text{tr}[R]}}[2E_c^{\top}(R)+e_Re_R^{\top}].\label{eq:E_inv}
\end{align}
Lastly, we re-state the matrices $E(R)$ and $E^{-1}(R)$ using the quaternion parameterization:
\begin{align}
\eta&\triangleq\cos\frac{\theta}{2},&\varepsilon&\triangleq\left(\sin\frac{\theta}{2}\right)\alpha.
\end{align}
In particular, we express the matrix $E(R)$ and its inverse as:
\begin{align}
E(R) 
&=\frac{1}{|\eta|}(\eta^2I+\eta\varepsilon^{\times}),&\eta&\neq0,\label{eq:E_quat}\\
E^{-1}(R)&=\frac{1}{|\eta|}\left(\eta^2I-\eta\varepsilon^{\times}+\varepsilon\varepsilon^{\top}\right),&\eta&\neq0.\label{eq:E_inv_quat}
\end{align}

\section{Problem Formulation}\label{sec:ProblemFormulation}
Let $R\in SO(3)$ denote the orientation of a body-fixed reference frame relative to an inertial reference frame, and let $\omega\in\mathbb{R}^3$ denote the body angular velocity expressed in body coordinates. Then, the attitude kinematics is given by:
\begin{align}
\dot{R}&=R\omega^{\times},&R(0)=R_{0}.\label{eq:Kinematics}
\end{align}
Suppose also that the true angular velocity $\omega$ is related to the measured angular velocity $\omega_{m}\in\mathbb{R}^3$ as:
\begin{equation}
\omega=\omega_{m}-\beta+B_1\delta_{1},\label{eq:MeasuredAngularVelocity}
\end{equation}
where $\beta\in\mathbb{R}^3$ is the measurement bias, $\delta_{1}\in\mathbb{R}^3$ is the process disturbance affecting the angular velocity measurement, and $B_1\in\mathbb{R}^{3\times3}$ is a filter tuning parameter. 
In addition, the bias term is modeled as:
\begin{align}
\dot{\beta} & =B_2\delta_{2}, & \beta(0)=\beta_{0},\label{eq:BiasModel}
\end{align}
where $\delta_{2}\in\mathbb{R}^3$ is the process disturbance affecting the bias, and $B_2\in\mathbb{R}^{3\times3}$ is also a filter tuning parameter. Lastly, we assume the availability of attitude measurements 
\begin{equation}
Y=R\epsilon,\label{eq:MeasuredAngle}
\end{equation}
where $\epsilon\in SO(3)$ represents the measurement error.

For the kinematic system \eqref{eq:Kinematics}-\eqref{eq:MeasuredAngle}, 
we consider the following estimator:
\begin{align}
\begin{bmatrix}\dot{\hat{R}}\\
\dot{\hat{\beta}}
\end{bmatrix}&=\begin{bmatrix}\hat{R}(\omega_{m}-\hat{\beta}+G_{q}\psi(\hat{R}^{\top}Y))^{\times}\\
G_{b}\psi(\hat{R}^{\top}Y)
\end{bmatrix},\label{eq:Estimator}
\end{align}
where $\psi$ denotes the error vector \eqref{eq:psi}, and $(G_{q},G_b)\in\mathbb{R}^{3\times3}$ represent filter gains which can either be fixed or time-varying. In particular, the error vector $\psi(\hat{R}^{\top}Y)$ represents the innovation term obtained using the estimated attitude $\hat{R}$ and the measured attitude $Y$. 

In order to formulate the main filtering problem addressed in this paper, we consider the following attitude and bias estimation errors:
\begin{align}
\tilde{R} & \triangleq\hat{R}^{\top}R, & \tilde{\beta}\triangleq\beta-\hat{\beta}.\label{eq:EstimationErrors}
\end{align}
Substituting the attitude kinematics \eqref{eq:Kinematics}-\eqref{eq:MeasuredAngularVelocity}, the bias model \eqref{eq:BiasModel}, and the estimator kinematics \eqref{eq:Estimator}, we obtain the estimation error system as:
\begin{subequations}
\begin{align}
\dot{\tilde{R}} & =\tilde{R}\omega_{m}^{\times}-\omega_{m}^{\times}\tilde{R}-\tilde{R}\beta^{\times}+\hat{\beta}^{\times}\tilde{R}+\tilde{R}(B_1\delta_{1})^{\times}\nonumber\\
&\quad\;-(G_{q}\psi(\tilde{R}\epsilon))^{\times}\tilde{R}\label{eq:Rtilde_dot} \\
\dot{\tilde{\beta}} & =B_2\delta_{2}-G_{b}\psi(\tilde{R}\epsilon)\label{eq:betatilde_dot}
\end{align} \label{eq:EstimationErrorSystem}
\end{subequations}
Next, we use the error vector \eqref{eq:eR} to define the following \emph{generalized} estimation error:
\begin{equation}
e_{R}(\tilde{R})=\frac{1}{\sqrt{1+\text{tr}[\tilde{R}]}}(\tilde{R}-\tilde{R}^{\top})^{\vee}.\label{eq:eR_Rtilde}
\end{equation}
Using the generalized estimation error and the bias error, we define the \textcolor{black}{filter error state} as:
\begin{equation}
z\triangleq\begin{bmatrix}e_{R}\\
\tilde{\beta}
\end{bmatrix}.\label{eq:z}
\end{equation}

\begin{proposition}\label{prop1}
Consider the system \eqref{eq:Kinematics}-\eqref{eq:MeasuredAngle} and the estimator \eqref{eq:Estimator}. Suppose that the generalized estimation error is given by \eqref{eq:eR_Rtilde}. Then, its time derivative can be expressed as:
\begin{equation}
\dot{e}_{R}=e_{R}^{\times}(\omega_{m}-\hat{\beta})+E(\tilde{R})(B_1\delta_{1}-\tilde{\beta})-E^{\top}(\tilde{R})G_{q}\psi(\tilde{R}\epsilon),\label{eq:eR_dot}
\end{equation}
where
\begin{equation}
E(\tilde{R})=\frac{1}{\sqrt{1+\text{tr}[\tilde{R}]}}\left(2E_c(\tilde{R})+\frac{1}{2}e_{R}e_{R}^{\top}\right),\label{eq:E_Rtilde}
\end{equation}
and
\begin{equation}
E_c(\tilde{R})=\frac{1}{2}\left(\text{tr}[\tilde{R}]I-\tilde{R}^{\top}\right).\label{eq:Ec_Rtilde}
\end{equation}
Furthermore, using \eqref{eq:betatilde_dot} and \eqref{eq:eR_dot}, we obtain the time derivative of the penalty variable \eqref{eq:z} as:
\begin{equation}
\dot{z}=Fz+\begin{bmatrix}E(\tilde{R})B_1\\
0
\end{bmatrix}\delta_{1}+\begin{bmatrix}0\\
B_2
\end{bmatrix}\delta_{2}-\begin{bmatrix}E^{\top}(\tilde{R})G_{q}\\
G_{b}
\end{bmatrix}\psi(\tilde{R}\epsilon),\label{eq:z_dot}
\end{equation}
where $F\in\mathbb{R}^{6\times6}$ denotes the following matrix:
\begin{equation}
F(\omega_m,\hat{\beta},R,\hat{R})\triangleq\begin{bmatrix}-\left(\omega_{m}-\hat{\beta}\right)^{\times} & -E(\tilde{R})\\
0 & 0
\end{bmatrix}.\label{eq:F}
\end{equation}
\end{proposition}

\begin{proof}
See Appendix \ref{sec:proof_prop1}.
\end{proof}

We consider the following candidate Lyapunov function:
\begin{equation}
V(z,t)=z^{\top}P^{-1}(t)z,\label{eq:StorageFunction}
\end{equation}
where $P(t)\in\mathbb{R}^{6\times6}$ is symmetric positive definite and has time-varying entries. The Lyapunov rate is:
\begin{equation}
\dot{V}=-z^{\top}P^{-1}\dot{P}P^{-1}z+2z^{\top}P^{-1}\dot{z},\label{eq:Vdot}
\end{equation}
where $\dot{z}$ is specified in \eqref{eq:z_dot}. We now state the main filtering problem addressed in this paper. Our goal is to obtain time-varying filter gains $(G_{q},G_{b})$ and a Riccati-type gain update equation such that 
the desired equilibrium $(\tilde{R},\tilde{\beta})=(I,0)$ of the estimation error system \eqref{eq:EstimationErrorSystem} is rendered almost globally uniformly asymptotically stable (AGUAS). To this end, we will establish that the proposed filter will, in the absence of process disturbances and measurement errors, that is, when $\delta_1=\delta_2=0$ and $\epsilon=I$, ensure that the Lyapunov rate \eqref{eq:Vdot} is negative definite for all initial conditions except those where $\text{tr}[\tilde{R}]=-1$.

\begin{remark}
The candidate Lyapunov function \eqref{eq:StorageFunction} is motivated by earlier work on nonlinear $H_{\infty}$ filtering for attitude and bias estimation using the quaternion representation \cite{markley1993h,markley1994deterministic}. In particular, the main difference between the Lyapunov function \eqref{eq:StorageFunction} and the storage function used in \cite{markley1993h,markley1994deterministic} is that the former employs the generalized estimation error \eqref{eq:eR_Rtilde} whereas the latter uses the vector part of the multiplicative quaternion error. 
\end{remark}


\section{Proposed Filter and Stability Analysis}\label{sec:GeneralizedSO3MEKF}

This section proposes a novel geometric nonlinear filter for attitude and bias estimation on $SO(3)$, using matrix measurements, and analyzes its stability. 
The proposed filter 
has the desirable property that, in the limit $\tilde{R}\rightarrow I$, it coincides with a variant of the multiplicative extended Kalman filter (MEKF), namely, the $SO(3)$-version with matrix measurements and biased velocity measurements. 

\subsection{Proposed Filter}\label{subsec:FilterGain}

We begin this part of the discussion by stating the filter gain and gain update equations for the $SO(3)$-MEKF:
\begin{subequations}
\begin{align}
\begin{bmatrix}G_{q}\\G_{b}\end{bmatrix} &= P\begin{bmatrix}I\\0\end{bmatrix}Q_3^{-1} \label{eq:FilterGain_Standard}, \\
 \dot{P} &= F_0P+PF_0^{\top}+\begin{bmatrix}Q_1&0\\0&Q_2\end{bmatrix}-P\begin{bmatrix}Q_3^{-1}&0\\0&0\end{bmatrix}P, \label{eq:GainUpdateEquation_MEKF_Standard} 
\end{align} \label{eq:SO3_MEKF}
\end{subequations}
where $F_0\in\mathbb{R}^{6\times6}$ denotes the following matrix:
\begin{equation}
F_0(\omega_m,\hat{\beta}) \triangleq \begin{bmatrix}-\left(\omega_{m}-\hat{\beta}\right)^{\times} & -I\\
0 & 0
\end{bmatrix}.\label{eq:F0}
\end{equation}
Furthermore, the filter tuning parameters $(Q_1,Q_2,Q_3)\in\mathbb{R}^{3\times3}$ are defined as:
\begin{align}
Q_1& \triangleq B_1B_1^{\top},& Q_2& \triangleq B_2B_2^{\top},&Q_3 \triangleq KK^{\top}.\label{eq:Q1Q2}
\end{align} 

Next, we state the condition on the filter gains $(G_q,G_b)$ which will play a crucial role in the stability analysis of the proposed filter. To this end, we consider the situation where the measurement error $\epsilon=I$, and modify the filter gain as:
\begin{equation}
\begin{bmatrix}E^{\top}(\tilde{R})G_{q}\\
G_{b}
\end{bmatrix}=\frac{1}{2}\left(\sqrt{1+\text{tr}[\tilde{R}]}\right)P\begin{bmatrix}I\\
0
\end{bmatrix}Q_3^{-1}\label{eq:MinimizingGain}
\end{equation}
From \eqref{eq:E_inv}, we recall that:
\begin{align*}
E^{-1}(\tilde{R})&=\frac{1}{\sqrt{1+\text{tr}[\tilde{R}]}}[2E_c^{\top}(\tilde{R})+e_Re_R^{\top}],&\text{tr}[\tilde{R}]\neq-1.\label{eq:E_inv_Rtilde}
\end{align*}
Taking the transpose of the matrix inverse and substituting \eqref{eq:Ec_Rtilde}, we obtain the \emph{ideal} filter gain as:
\begin{equation}
\begin{bmatrix} G_{q} \\ G_{b} \end{bmatrix} =\frac{1}{2} \begin{bmatrix} \left(\text{tr}[\tilde{R}]I-\tilde{R}+e_Re_R^{\top}\right) \\
\left(\sqrt{1+\text{tr}[\tilde{R}]}\right)I \end{bmatrix} P \begin{bmatrix} I \\ 0 \end{bmatrix} Q_3^{-1}.\label{eq:FilterGain_Ideal}
\end{equation}
Lastly, we modify the gain update equation as follows:
\begin{equation}
\begin{split}
\dot{P}&=FP+PF^{\top}+\begin{bmatrix}E(\tilde{R})Q_1E^{\top}(\tilde{R})&0\\0&Q_2\end{bmatrix}\\
&\quad-\frac{1}{4}(1+\text{tr}[\tilde{R}])P\begin{bmatrix}I\\0\end{bmatrix}Q_3^{-1}\begin{bmatrix}I&0\end{bmatrix}P.\label{eq:GainUpdateEquation_MEKF_Ideal}
\end{split}
\end{equation}
where $F\in\mathbb{R}^{6\times6}$ is defined in \eqref{eq:F}. 

\begin{remark}
In the limit $\tilde{R}\rightarrow I$, we note that:
\begin{align*}
e_R&\rightarrow0,&E(\tilde{R})&\rightarrow I,
\end{align*}
where $e_R$ denotes the generalized estimation error \eqref{eq:eR_Rtilde}, and the matrix $E(\tilde{R})$ is defined in \eqref{eq:E_Rtilde}.  
Thus, in the limit $\tilde{R}\rightarrow I$, the filter gain \eqref{eq:FilterGain_Ideal} and the gain update equation \eqref{eq:GainUpdateEquation_MEKF_Ideal} reduce to those used in the $SO(3)$-MEKF.
\end{remark}


The filter gain \eqref{eq:FilterGain_Ideal} and the gain update equation \eqref{eq:GainUpdateEquation_MEKF_Ideal}, of which \eqref{eq:FilterGain_Standard}-\eqref{eq:GainUpdateEquation_MEKF_Standard} are special cases, represent an \emph{ideal} situation where the true estimation error $\tilde{R}=\hat{R}^{\top}R$ is known. 
Since the true attitude $R$ is unknown, we replace it by its measured value $Y$, and
obtain the following filter:
\begin{subequations}
\begin{align}
\begin{bmatrix} G_{q} \\ G_{b} \end{bmatrix} &= \frac{1}{2} \begin{bmatrix} \left(\text{tr}[M]I-M+e_Ye_Y^{\top}\right) \\
\left(\sqrt{1+\text{tr}[M]}\right) \end{bmatrix} P \begin{bmatrix} I \\ 0 \end{bmatrix} Q_3^{-1},\label{eq:FilterGain_Implementable} \\
 \dot{P} &= F_YP+PF_Y^{\top}+\begin{bmatrix}E(M)Q_1E^{\top}(M)&0\\0&Q_2\end{bmatrix} \nonumber \\
 &\quad-\frac{1}{4}(1+\text{tr}[M])P\begin{bmatrix}I\\0\end{bmatrix}Q_3^{-1}\begin{bmatrix}I&0\end{bmatrix}P\label{eq:GainUpdateEquation_MEKF_SO3},
\end{align} \label{eq:Generalized_MEKF_SO3}
\end{subequations}
where $M=\hat{R}^{\top}Y\in SO(3)$, and 
\begin{equation}
\begin{split}
e_Y(M) &= \frac{1}{\sqrt{1+\text{tr}[M]}}\left(M-M^{\top}\right)^{\vee}, \\ 
E(M) &= \frac{1}{\sqrt{1+\text{tr}[M]}}\left[\frac{1}{2}(1+\text{tr}[M])I+\psi^{\times}(M)\right], \\ 
F_Y &= \begin{bmatrix}-\left(\omega_{m}-\hat{\beta}\right)^{\times} & -E(M)\\
0 & 0
\end{bmatrix}. 
\end{split} \label{eq:eY_E_Fm}
\end{equation}


\subsection{Filter Stability}

In order to analyze the closed-loop stability of the proposed $SO(3)$-MEKF \eqref{eq:Generalized_MEKF_SO3}-\eqref{eq:eY_E_Fm}, we make the following assumption for the filter gain $P(t)$.

\begin{assumption} \label{assumption:Pt}
\textcolor{black}{There exist scalars $\underline{p},\overline{p}>0$ such that $\underline{p}I\leq P(t)\leq\overline{p}I$.}
\end{assumption}

\begin{theorem}
Consider the kinematic system \eqref{eq:Kinematics}-\eqref{eq:MeasuredAngle}, the estimator \eqref{eq:Estimator}, the candidate Lyapunov function \eqref{eq:StorageFunction}, and the proposed filter \eqref{eq:Generalized_MEKF_SO3}-\eqref{eq:eY_E_Fm}. 
Suppose that Assumption 1 holds, the process disturbance $\delta_1=\delta_2=0$, the measurement error $\epsilon=I$, and the initial estimation error $\tilde{R}(0)$ satisfies $\text{tr}[\tilde{R}]\neq-1$. Then, the Lyapunov rate \eqref{eq:Vdot} is \textcolor{black}{negative definite}, and the desired equilibrium $z=(e_R,\tilde{\beta})=(0,0)$ of the estimation error system \eqref{eq:z_dot} is almost globally uniformly asymptotically stable (AGUAS).
\end{theorem}

\begin{proof}
We begin the proof by substituting $\delta_1=\delta_2=0$ and $\epsilon=I$ in \eqref{eq:z_dot} to obtain:
\begin{equation}
\dot{z}=Fz-\begin{bmatrix}E^{\top}(\tilde{R})G_{q}\\G_{b}\end{bmatrix}\psi(\tilde{R}).\label{eq:z_dot_Ideal}
\end{equation}
We note that for $\epsilon=I$, the filter gain and gain update equation are given by \eqref{eq:FilterGain_Ideal} and \eqref{eq:GainUpdateEquation_MEKF_Ideal}, respectively. Furthermore, since the former follows directly from the gain condition \eqref{eq:MinimizingGain}, it satisfies this condition. In addition, from \eqref{eq:psi} and \eqref{eq:eR_Rtilde}, we observe that:
\begin{equation}
2\psi(\tilde{R}) = (\tilde{R}-\tilde{R}^{\top})^{\vee} = \left(\sqrt{1+\text{tr}[\tilde{R}]}\right)e_R.\label{eq:RRTvee}
\end{equation}
Consequently, using \eqref{eq:MinimizingGain} and \eqref{eq:RRTvee}, we simplify the second term in \eqref{eq:z_dot_Ideal} as follows:
\begin{align*}
\begin{bmatrix}E^{\top}(\tilde{R})G_{q}\\G_{b}\end{bmatrix}\psi(\tilde{R})&=\frac{1}{2}\left(\sqrt{1+\text{tr}[\tilde{R}]}\right)P\begin{bmatrix}I\\0\end{bmatrix}Q_3^{-1}\psi(\tilde{R})\\
&=\frac{1}{4}\left(1+\text{tr}[\tilde{R}]\right)P\begin{bmatrix}I\\0\end{bmatrix}Q_3^{-1}\begin{bmatrix}I&0\end{bmatrix}z.
\end{align*}
Substituting this expression in \eqref{eq:z_dot_Ideal}, and the result in \eqref{eq:Vdot}, we express the Lyapunov rate as:
\begin{equation}
\begin{split}
\dot{V}
&=-z^{\top}P^{-1}\dot{P}P^{-1}z+z^{\top}(P^{-1}F+F^{\top}P^{-1})z \\ 
&\quad-\frac{1}{2}\left(1+\text{tr}[\tilde{R}]\right)z^{\top}\begin{bmatrix}I\\0\end{bmatrix}Q_3^{-1}\begin{bmatrix}I&0\end{bmatrix}z
\end{split} \label{eq:Vdot_Lyapunov}
\end{equation}
Next, we substitute the gain update equation \eqref{eq:GainUpdateEquation_MEKF_Ideal} and simplify the first term in \eqref{eq:Vdot_Lyapunov} as:
\begin{equation*}
\begin{split}
-z^{\top}P^{-1}\dot{P}P^{-1}z=&-z^{\top}P^{-1}(FP+PF^{\top})P^{-1}z\\
&-z^{\top}P^{-1}\begin{bmatrix}E(\tilde{R})Q_1E^{\top}(\tilde{R})&0\\0&Q_2\end{bmatrix}P^{-1}z\\
&+\frac{1}{4}(1+\text{tr}[\tilde{R}])z^{\top}\begin{bmatrix}I\\0\end{bmatrix}Q_3^{-1}\begin{bmatrix}I&0\end{bmatrix}z
\end{split}
\end{equation*}
As a result, the Lyapunov rate simplifies to:
\begin{equation}
\begin{split}
\dot{V}=&-z^{\top}P^{-1}\begin{bmatrix}E(\tilde{R})Q_1E^{\top}(\tilde{R})&0\\0&Q_2\end{bmatrix}P^{-1}z\\
&-\frac{1}{4}\left(1+\text{tr}[\tilde{R}]\right)z^{\top}\begin{bmatrix}I\\0\end{bmatrix}Q_3^{-1}\begin{bmatrix}I&0\end{bmatrix}z
\end{split}
\end{equation}
Suppose:
\begin{equation}
Q_G(\tilde{R}):=\begin{bmatrix}E(\tilde{R})Q_1E^{\top}(\tilde{R})&0\\0&Q_2\end{bmatrix}.
\end{equation}
Then, we can bound the Lyapunov rate as:
\begin{equation}
\dot{V}\leq-z^{\top}P^{-1}Q_G(\tilde{R})P^{-1}z.\label{eq:Vdot_B}
\end{equation}
Since $Q_1$ is positive definite, it follows that the product $E(\tilde{R})Q_1E^{\top}(\tilde{R})$ is positive semi-definite. Furthermore, since the determinant of $E(\tilde{R})$ is
\begin{equation*}
\det(E(\tilde{R}))=\sqrt{\frac{1+\text{tr}[\tilde{R}]}{4}},
\end{equation*}
it follows that $\tilde{R}$ has rank $3$ for all estimation errors $\tilde{R}$ except those corresponding to $\text{tr}\;[\tilde{R}]=-1$, that is, estimation errors in which the angle of rotation equals $\pi$ radians. Excluding these points from the analysis, we note that for the rest, the product $E(\tilde{R})Q_1E^{\top}(\tilde{R})$ is positive definite. Since $Q_2$ is also positive definite, it follows that
\begin{equation}
Q_G(\tilde{R})>0,\quad\text{tr}[\tilde{R}]\neq-1.\label{eq:QG_PosDef}
\end{equation}
\textcolor{black}{Combining \eqref{eq:Vdot_B} and \eqref{eq:QG_PosDef}, we observe that for attitude estimation errors in which $\text{tr}[\tilde{R}]\neq-1$, the Lyapunov rate is bounded by a continuous negative definite function. 
%
%
From Assumption 1, it follows that:
\begin{equation}
\underline{p} \|z\|^2 \leq V(z,t) \leq \overline{p} \|z\|^2. \label{eq:V_LBUB}
\end{equation}
Consequently, from \eqref{eq:Vdot_B}-\eqref{eq:V_LBUB}}, we observe that the quadratic Lyapunov function $V$ and its time derivative $\dot{V}$ satisfy the conditions stipulated in \cite[Theorem 4.9]{Khalil2002}. This allows us to conclude that the estimation error $z$
converges to zero almost globally uniformly asymptotically, that is, for almost all initial conditions except those corresponding to initial attitude estimation errors with $\text{tr}[\tilde{R}]=-1$.
\end{proof}

\begin{remark}
\textcolor{black}{The above result establishes the AGUAS property for the proposed Generalized $SO(3)$-MEKF using biased velocity measurements and matrix attitude measurements. In comparison, the traditional $SO(3)$-MEKF has, to the best of our knowledge, only local stability guarantees. Global exponential stability has been established for the multiplicative exogenous Kalman filter (MXKF), proposed in \cite{stovner2018attitude}. The MXKF consists of the cascade connection of a fixed-gain observer and the multiplicative extended Kalman filter. On the other hand, the filter proposed in this paper is a single-stage filter which generalizes the $SO(3)$-MEKF.}
\end{remark}

\begin{remark}
\textcolor{black}{
Most of the studies on the stability of the extended Kalman filter (EKF) have been conducted using Assumption \ref{assumption:Pt}. These include the pioneering results reported in \cite{deyst1968conditions,baras1988dynamic,song_grizzle1995}, as well as the contraction-based approach followed by Bonnabel and Slotine \cite{bonnabel2014contraction}. The work by Krener \cite{krener2002mathematical} is among the few papers to have studied EKF stability without invoking Assumption \ref{assumption:Pt}. However, in all these papers, only local convergence results have been obtained. On the other hand, in this paper, we have used Assumption \ref{assumption:Pt} to obtain a novel geometric nonlinear filter which almost globally uniformly asymptotically stabilizes the origin of the estimation error system \eqref{eq:z_dot}. Furthermore, we have shown that the proposed filter can be interpreted as a more general form of the $SO(3)$-MEKF.
}
\end{remark}

\subsection{Quaternion Implementation}

In their seminal work \cite{mahony2008nonlinear}, Mahony et al. use the rotation matrix parameterization for filter analysis and recommend the use of the quaternion representation for filter implementation. Following in their footsteps, we now obtain the quaternion counterpart of the ideal filter gain \eqref{eq:FilterGain_Ideal} and the proposed filter \eqref{eq:Generalized_MEKF_SO3}-\eqref{eq:eY_E_Fm}. To this end, we restate the gain condition \eqref{eq:MinimizingGain} as:
\begin{equation}
\begin{bmatrix}E^{\top}(\tilde{R})G_{q}\\
G_{b}
\end{bmatrix}= |\tilde{\eta}| P \begin{bmatrix}I\\
0
\end{bmatrix}Q_3^{-1}, \label{eq:MinimizingGain_Partitioned_Quat}
\end{equation}
where the matrix $E(\tilde{R})$ is obtained using \eqref{eq:E_quat}, that is,
\[
E(\tilde{R})=\frac{1}{|\tilde{\eta}|}(\tilde{\eta}^2I+\tilde{\eta}\tilde{\varepsilon}^{\times}),
\]
and $(\tilde{\eta},\tilde{\varepsilon})$ denote either of the two quaternions associated with the attitude estimation error $\tilde{R}$. From \eqref{eq:E_inv_quat}, we recall that:
\begin{align}
E^{-1}(\tilde{R})&=\frac{1}{|\tilde{\eta}|}\left(\tilde{\eta}^2I-\tilde{\eta}\tilde{\varepsilon}^{\times}+\tilde{\varepsilon}\tilde{\varepsilon}^{\top}\right),&\tilde{\eta}&\neq0.
\end{align}
Consequently, we obtain the ideal filter gain from \eqref{eq:MinimizingGain_Partitioned_Quat} as:
\begin{equation}
\begin{bmatrix}G_{q}\\
G_{b}
\end{bmatrix}=\begin{bmatrix}(\tilde{\eta}^2I+\tilde{\eta}\tilde{\varepsilon}^{\times}+\tilde{\varepsilon}\tilde{\varepsilon}^{\top})\\
|\tilde{\eta}|I
\end{bmatrix} P \begin{bmatrix} I \\ 0 \end{bmatrix} Q_3^{-1}\label{eq:FilterGain_Ideal_Quat}
\end{equation}
As before, we obtain an implementable form of the ideal filter gain and gain update equation by replacing the true attitude $R$ by its measured value $Y$. Consequently, we obtain the quaternion counterpart of the proposed filter \eqref{eq:Generalized_MEKF_SO3}-\eqref{eq:eY_E_Fm} as:
\begin{subequations}
\begin{align}
\begin{bmatrix}G_{q}\\
G_{b}
\end{bmatrix} &= \begin{bmatrix}(\eta^2I+\eta\epsilon^{\times}+\epsilon\epsilon^{\top})\\
|\eta|I,
\end{bmatrix}P\begin{bmatrix}I\\0\end{bmatrix}Q_3^{-1},\label{eq:FilterGain_Implementable_Quat} \\
 \dot{P} &= F_qP+PF_q^{\top}+\begin{bmatrix}[\eta I+\varepsilon^{\times}]Q_1[\eta I-\varepsilon^{\times}]&0\\0&Q_2\end{bmatrix}\\
 &\quad-P\begin{bmatrix}\eta^2Q_3^{-1}&0\\0&0\end{bmatrix}P+\frac{1}{\gamma^{2}}P^2,\label{eq:GainUpdateEquation_Hinfty_Quat}
\end{align} \label{eq:S3_Generalized_MEKF}
\end{subequations}
where $(\eta,\epsilon)\in\mathbb{S}^3$ denote the components of either of the two quaternions associated with the measured estimation error $M=\hat{R}^{\top}Y\in SO(3)$, the matrix $F_q\in\mathbb{R}^{6\times6}$ is defined given as
\begin{equation}
F_q(\omega_m,\hat{\beta},\eta,\varepsilon)\triangleq\begin{bmatrix}-\left(\omega_{m}-\hat{\beta}\right)^{\times} & -\text{sgn}(\eta)[\eta I+\varepsilon^{\times}]\\
0 & 0
\end{bmatrix},\label{eq:Fq}
\end{equation}
and $\text{sgn}(\cdot)$ denotes the signum function. Table \ref{tab:MEKF} summarizes the filter equations for the different MEKF variants.

\begin{table}[tbh]
\centering
\caption{Equations for the different MEKF variants.\label{tab:MEKF}}
\begin{tabular}{ccc}
\hline
Filter & Filter gain \\\hline
Standard $SO(3)$-MEKF & \eqref{eq:SO3_MEKF}-\eqref{eq:F0}\\
Ideal $SO(3)$-MEKF & \eqref{eq:FilterGain_Ideal}-\eqref{eq:GainUpdateEquation_MEKF_Ideal}\\
Proposed $SO(3)$-MEKF & \eqref{eq:Generalized_MEKF_SO3}-\eqref{eq:eY_E_Fm}\\
Proposed $S^3$-MEKF & \eqref{eq:S3_Generalized_MEKF}-\eqref{eq:Fq}\\
\hline
\end{tabular}
\end{table}


\section{Simulation Results}\label{sec:SimulationResults}
This section presents simulation results comparing the performance of the proposed filter \eqref{eq:Generalized_MEKF_SO3}-\eqref{eq:eY_E_Fm} with that of the traditional $SO(3)$-MEKF (using matrix measurements) and the nonlinear complementary filter \cite{mahony2008nonlinear}. All three filters are implemented using the quaternion parameterization.

\subsection{Problem Setup and Gain Selection}

The true attitude trajectory is obtained as:
\begin{equation}
\dot{R}=R\omega^{\times},\quad R(0)=R_0,\label{eq:Kinematics_Simulation}
\end{equation}
where $R_0\in SO(3)$ denotes the initial attitude, and the true angular velocity is:
\begin{equation}
\omega=\begin{bmatrix}\cos3t,&0.1\sin2t,&-\cos t\end{bmatrix}^{\top}.
\end{equation}
The measured angular velocity is obtained as:
\begin{equation}
\omega_m=\omega+\beta+\delta_1,
\end{equation}
where $\beta\in\mathbb{R}^3$ is the unknown gyro bias, and $\delta_1\in\mathbb{R}^3$ is a zero-mean white Gaussian noise with standard deviation $\sigma_{\omega}=\sqrt{\pi/12}\;\text{rad/s}$. The time evolution of the bias is modeled as:
\begin{align}
\dot{\beta}&=\delta_2,&\beta(0)&=\beta_0,\label{eq:BiasModel_Simulation}
\end{align}
where $\delta_2\in\mathbb{R}^3$ is a zero-mean white Gaussian noise with standard deviation $\sigma_{\beta}=\sqrt{\pi/180}\;{\text{rad}/\text{s}^2}$.
The attitude trajectory \eqref{eq:Kinematics_Simulation} is initialized at the rotation matrix $R_0$ corresponding to the initial Euler angles $\begin{bmatrix}\pi&-\pi/2&\pi/2\end{bmatrix}^{\top}$, and the bias model \eqref{eq:BiasModel_Simulation} is initialized at $\beta_0=\pi/4\;\text{rad/s}$. In order to simulate matrix measurements for the attitude, we consider the following body-frame vectors:
\begin{align}
y_1&=R^{\top}e_1+\epsilon_1,&y_2&=R^{\top}e_2+\epsilon_2,\label{eq:AttitudeMeasurement}
\end{align}
where $e_1=\begin{bmatrix}1&0&0\end{bmatrix}^{\top}$, $e_2=\begin{bmatrix}0&1&0\end{bmatrix}^{\top}$, and $\epsilon_i\in\mathbb{R}^3$ denotes zero-mean white Gaussian noise with standard deviation $\sigma_{\epsilon}=\sqrt{\pi/12}$. We use the the vector measurements $(y_1,y_2)$, in conjunction with the TRIAD algorithm \cite{black1964passive}, \cite[Section 4.1]{hashim2020attitude}, to simulate attitude measurements of the form $Y=R\epsilon$.


For each of the filters under consideration, we select its gains in such a way that the three filters have comparable steady-state performance. To this end, we use a simplified form of the MEKF gain update equation \eqref{eq:GainUpdateEquation_MEKF_Standard} to select the inital gain $P(0)$. In particular, we observe that for the angle and bias estimation problem on the unit circle using angular measurements, the filter gain \eqref{eq:FilterGain_Standard} for the MEKF can be simplified as:
\begin{equation}
\begin{bmatrix}G_q\\G_b\end{bmatrix}=\frac{1}{q_3}\begin{bmatrix}p_a&p_b\\p_b&p_c\end{bmatrix}\begin{bmatrix}1\\0\end{bmatrix},\label{eq:FilterGain_Standard_S1}
\end{equation}
and the gain update equation \eqref{eq:GainUpdateEquation_MEKF_Standard} as:
\begin{align*}
\dot{p}_a&=-2p_b+q_1-\frac{1}{q_3}p_a^2,\\
\dot{p}_b&=-p_c-\frac{1}{q_3}p_ap_b,\\
\dot{p}_c&=q_2-\frac{1}{q_3}p_b^2,
\end{align*}
Consequently, we obtain the following expressions for the steady-state values of the gains of the $S^1$-MEKF:
\begin{subequations}
\begin{align}
\dot{p}_c&=0\implies p_b^*=-\sqrt{q_2q_3},\\
\dot{p}_a&=0\implies p_a^*=\sqrt{q_3(q_1-2p_b^*)},\\
\dot{p}_b&=0\implies p_c^*=-\frac{1}{q_3}p_a^*p_b^*.
\end{align}\label{eq:pSS}
\end{subequations}
Next, we select $(q_1,q_2,q_3)$ as follows:
\begin{align*}
q_3&=(0.8\sigma_{\epsilon})^2,&q_2&=k_I^2q_3,&q_1&=q_3(k_P^2-2k_I),
\end{align*}
where $\sigma_{\epsilon}=\sqrt{\pi/12}$ is the standard deviation for the measurement error in \eqref{eq:AttitudeMeasurement}, and $(k_P,k_I)=(5.9126,1.7738)$ are the gains selected for the nonlinear complementary filter. This choice yields:
\begin{align*}
q_1&=5.263,&q_2&=0.5272,&q_3&=0.1676.
\end{align*}
Lastly, we substitute these values in \eqref{eq:pSS} and initialize the filter gains for the $SO(3)$-MEKF and the proposed $SO(3)$-MEKF as
\begin{equation}
P(0)=3\begin{bmatrix}p_a^*I&p_b^*I\\p_b^*I&p_c^*I\end{bmatrix}.\label{eq:P0}
\end{equation}
In this way, we ensure that the three filters under consideration have 
comparable steady-state performance.


\subsection{Results}

\begin{figure}
\centering
\includegraphics[width=0.99\columnwidth]{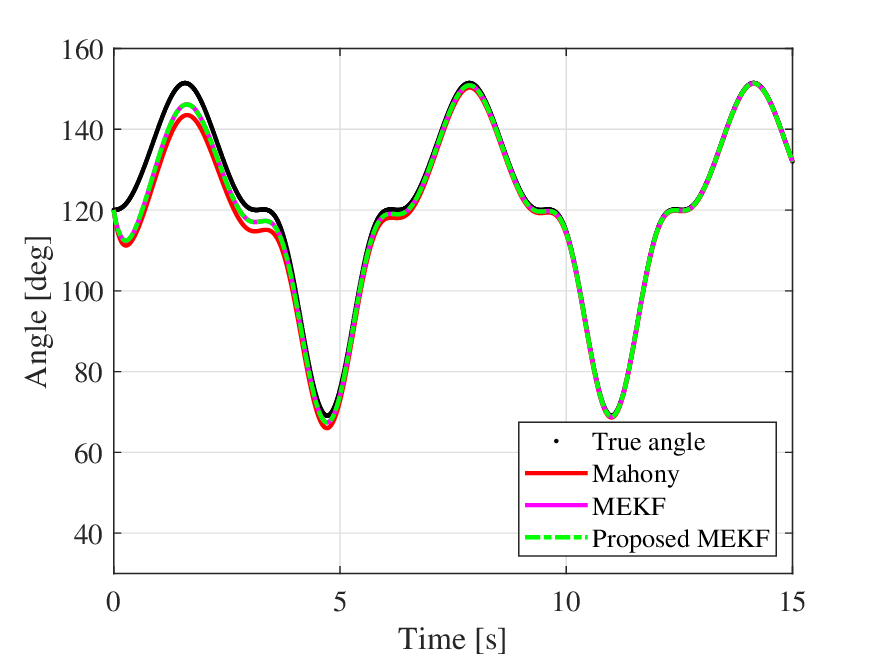}\caption{The true attitude trajectory and its estimates obtained using noise-free measurements\label{fig:Trajs_Angle_Ideal}}
\end{figure}

\begin{figure}
\centering
\includegraphics[width=0.99\columnwidth]{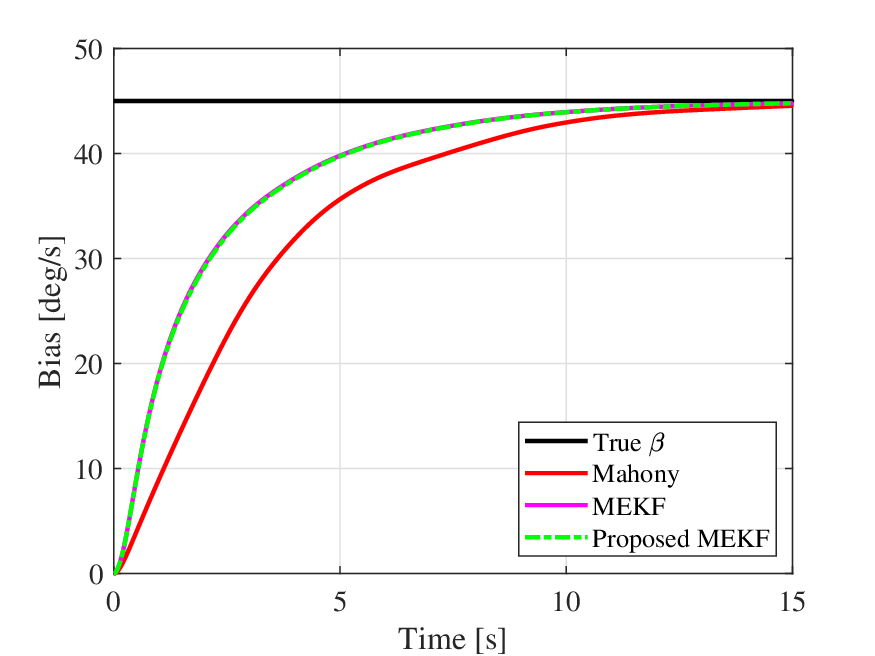}\caption{Bias estimates obtained using noise-free measurements\label{fig:Trajs_Bias_Ideal}}
\end{figure}

For our first simulation run, we consider the problem of estimating the unknown attitude and bias using noise-free measurements, that is, with the various process disturbances and measurement errors set to zero. Figure \ref{fig:Trajs_Angle_Ideal} plots the angle of rotation associated with the true attitude trajectory and its estimates obtained using different filters. Similarly, Fig. \ref{fig:Trajs_Bias_Ideal} plots the true (unknown) bias and its estimates.
These figures suggest that the MEKF and the proposed generalized MEKF have similar performance.
Next, we compare the filters in the presence of process disturbances and measurement errors.
In particular, we run $50$ simulations for each filter, and tabulate the RMS values of the angle of rotation and the bias error associated with the attitude and bias estimates, respectively. These values are reported in Tables \ref{tab:AttEstErrors-RMS_Table} and \ref{tab:BiasEstErrors-RMS_Table}. They indicate that the MEKF has the best steady-state performance. However,
in the case of attitude estimation, 
\textcolor{black}{the proposed filter has better transient performance than the MEKF and the nonlinear complementary filter.} 


\begin{figure}
\centering
\includegraphics[width=0.99\columnwidth]{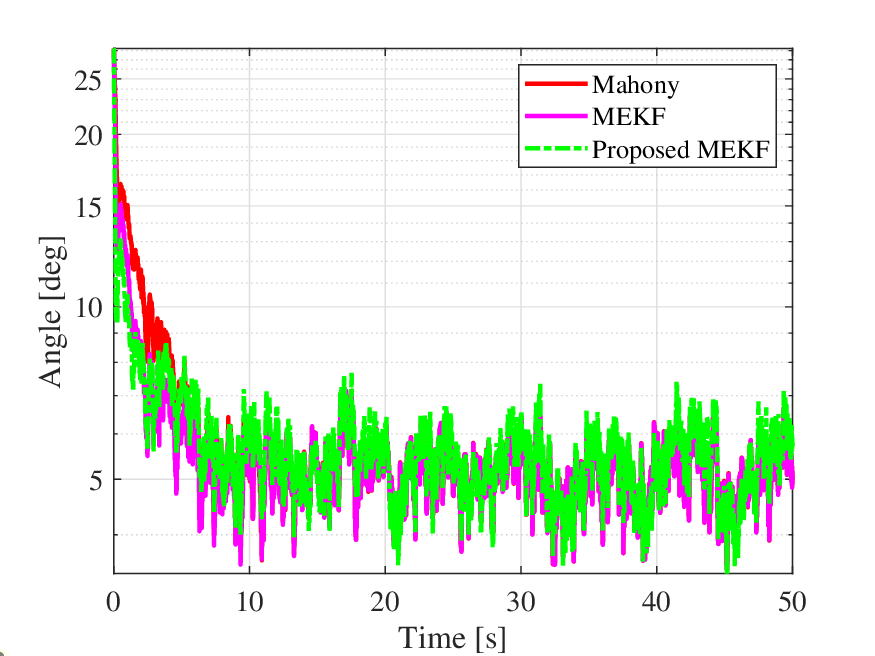}\caption{Mean RMS errors during attitude estimation\label{fig:Error_Angles_Noisy}}
\end{figure}




\begin{table}
\centering
\caption{Mean RMS errors (deg) for attitude estimation\label{tab:AttEstErrors-RMS_Table}}
\begin{tabular}{ccc} \hline
Filter & Transient & Steady state\tabularnewline 
\hline
TRIAD & 26.5455 & 26.9509\\
Mahony & 11.0330 & 5.1822\\
MEKF & 8.8719 & 5.1380\\
Proposed MEKF & 8.5026 & 5.3932\\
\hline
\end{tabular}
\end{table}

\begin{table}
\centering
\caption{Mean RMS errors (deg/s) for bias estimation\label{tab:BiasEstErrors-RMS_Table}}
\begin{tabular}{ccc} \hline
Filter & Transient & Steady state\tabularnewline 
\hline
Mahony & 25.4237 & 2.7408\\
MEKF & 18.9933 & 2.4334\\
Proposed MEKF & 20.8840 & 2.8714\\
\hline
\end{tabular}
\end{table}

\section{Conclusion}\label{sec:Conclusion}
This paper proposed the generalized $SO(3)$-MEKF for attitude and bias estimation on $SO(3)$. 
The proposed filter almost globally uniformly asymptotically stabilizes the desired equilibrium of the closed-loop estimation error system. Moreover, the traditional $SO(3)$-MEKF, with matrix measurements, is a special case of the proposed filter.

\appendices

\section{Supporting Results}

\subsection{Proof of Proposition \ref{prop1}}\label{sec:proof_prop1}

\begin{proof}
Applying the chain rule, we can express the time derivative of the generalized estimation error \eqref{eq:eR_Rtilde} as:
\begin{equation}
\dot{e}_{R}=\frac{1}{\sqrt{1+\text{tr}[\tilde{R}]}}(\dot{\tilde{R}}-\dot{\tilde{R}}^{\top})^{\vee}-\frac{\text{tr}[\dot{\tilde{R}}]}{2(1+\text{tr}[\tilde{R}])}e_R\label{eq:eRdot_Appendix}
\end{equation}
In order to obtain the rate of change of the vector $(\tilde{R}-\tilde{R}^{\top})^{\vee}$, we substitute $\dot{\tilde{R}}$ from \eqref{eq:Rtilde_dot} and proceed as follows:
\begin{align*}
\dot{\tilde{R}}-\dot{\tilde{R}}^{\top}
&=\tilde{R}\omega_{m}^{\times}-\omega_{m}^{\times}\tilde{R}-\tilde{R}\beta^{\times}+\hat{\beta}^{\times}\tilde{R}+\tilde{R}(B_1\delta_{1})^{\times}\\
&\quad-(G_{q}\psi(\tilde{R}\epsilon))^{\times}\tilde{R}+\omega_{m}^{\times}\tilde{R}^{\top}-\tilde{R}^{\top}\omega_{m}^{\times}-\beta^{\times}\tilde{R}^{\top}\\
&\quad+\tilde{R}^{\top}\hat{\beta}^{\times}+(B_1\delta_{1})^{\times}\tilde{R}^{\top}-\tilde{R}^{\top}(G_{q}\psi(\tilde{R}\epsilon))^{\times}\\
&=(\tilde{R}\omega_{m}^{\times}+\omega_{m}^{\times}\tilde{R}^{\top})-(\omega_{m}^{\times}\tilde{R}+\tilde{R}^{\top}\omega_{m}^{\times})\\
&\quad+(\tilde{R}(B_1\delta_{1})^{\times}+(B_1\delta_{1})^{\times}\tilde{R}^{\top})\\
&\quad-(\tilde{R}\beta^{\times}+\beta^{\times}\tilde{R}^{\top})+(\hat{\beta}^{\times}\tilde{R}+\tilde{R}^{\top}\hat{\beta}^{\times})\\
&\quad-((G_{q}\psi(\tilde{R}\epsilon))^{\times}\tilde{R}+\tilde{R}^{\top}(G_{q}\psi(\tilde{R}\epsilon))^{\times})
\end{align*}
Using the following identity,
\[
x^{\times}A+A^{\top}x^{\times}=\{(\text{tr}[A]I-A)x\}^{\times},\quad x\in\mathbb{R}^3,A\in\mathbb{R}^{3\times3},
\]
we re-state the preceding expression as:
\begin{align}
\frac{1}{2}(\dot{\tilde{R}}-\dot{\tilde{R}}^{\top})&=(E_c(\tilde{R})\omega_{m})^{\times}-(E_c^{\top}(\tilde{R})\omega_{m})^{\times}\nonumber\\
&\quad-(E_c(\tilde{R})\beta)^{\times}+(E_c^{\top}(\tilde{R})\hat{\beta})^{\times}\nonumber\\
&\quad+(E_c(\tilde{R})B_1\delta_{1})^{\times}-(E_c^{\top}(\tilde{R})G_{q}\psi(\tilde{R}\epsilon))^{\times},\label{eq:Rtilde-RtildeT_B}
\end{align}
where the mapping $E_c(\tilde{R})$ is defined in \eqref{eq:Ec_Rtilde}. For the first two terms on the right-hand side of \eqref{eq:Rtilde-RtildeT_B}, we observe that:
\[
2(E_c-E_c^{\top})=(\text{tr}[\tilde{R}]I-\tilde{R}^{\top})-(\text{tr}[\tilde{R}]I-\tilde{R})=\tilde{R}-\tilde{R}^{\top}.
\]
Similarly, we re-state the third and fourth terms as:
\begin{align*}
2(E_c\beta-E_c^{\top}\hat{\beta})&=2E_c\beta-(\text{tr}[\tilde{R}]I-\tilde{R}+\tilde{R}^{\top}-\tilde{R}^{\top})\hat{\beta}\\
&=2E_c(\beta-\hat{\beta})-(\tilde{R}^{\top}-\tilde{R})\hat{\beta}\\
&=2E_c\tilde{\beta}+(\tilde{R}-\tilde{R}^{\top})\hat{\beta}
\end{align*}
Consequently, we can re-state \eqref{eq:Rtilde-RtildeT_B} as:
\begin{align*}
\frac{1}{2}(\dot{\tilde{R}}-\dot{\tilde{R}}^{\top})&=\frac{1}{2}\{(\tilde{R}-\tilde{R}^{\top})(\omega_{m}-\hat{\beta})\}^{\times}-(E_c\tilde{\beta})^{\times}\\
&\quad+(E_cB_1\delta_{1})^{\times}-(E_c^{\top}G_{q}\psi(\tilde{R}\epsilon))^{\times}
\end{align*}
Removing the cross operator, it follows that:
\begin{align}
(\dot{\tilde{R}}-\dot{\tilde{R}}^{\top})^{\vee} & =(\tilde{R}-\tilde{R}^{\top})(\omega_{m}-\hat{\beta})+2E_c(B_1\delta_{1}-\tilde{\beta})\nonumber\\
&\quad-2E_c^{\top}G_{q}\psi(\tilde{R}\epsilon)\label{eq:Rtilde-RtildeT}
\end{align}
Substituting \eqref{eq:Rtilde-RtildeT} in \eqref{eq:eRdot_Appendix}, we express the time derivative of the generalized estimation error $e_R$ as follows:
\begin{align*}
\dot{e}_{R}&=\frac{1}{\sqrt{1+\text{tr}[\tilde{R}]}}\left[(\tilde{R}-\tilde{R}^{\top})(\omega_{m}-\hat{\beta})+2E_c(B_1\delta_{1}-\tilde{\beta})\right]\\
&\quad-\frac{2}{\sqrt{1+\text{tr}[\tilde{R}]}}E_c^{\top}G_{q}\psi(\tilde{R}\epsilon)-\frac{\text{tr}[\dot{\tilde{R}}]}{2(1+\text{tr}[\tilde{R}])}e_R\\
&=e_R^{\times}(\omega_{m}-\hat{\beta})+\frac{2}{\sqrt{1+\text{tr}[\tilde{R}]}}E_c(B_1\delta_{1}-\tilde{\beta})\\
&\quad-\frac{2}{\sqrt{1+\text{tr}[\tilde{R}]}}E_c^{\top}G_{q}\psi(\tilde{R}\epsilon)-\frac{\text{tr}[\dot{\tilde{R}}]}{2(1+\text{tr}[\tilde{R}])}e_R
\end{align*}
Applying the trace operator in \eqref{eq:Rtilde_dot}, we observe that:
\begin{align}
\text{tr}[\dot{\tilde{R}}] &=\text{tr}[\tilde{R}\omega_{m}^{\times}]-\text{tr}[\omega_{m}^{\times}\tilde{R}]-\text{tr}[\tilde{R}\beta^{\times}]+\text{tr}[\hat{\beta}^{\times}\tilde{R}]\nonumber\\
 &\quad+\text{tr}[\tilde{R}(B_1\delta_{1})^{\times}]-\text{tr}[(G_{q}\psi(\tilde{R}\epsilon))^{\times}\tilde{R}]\nonumber\\
 &=\text{tr}[\tilde{R}\omega_{m}^{\times}]-\text{tr}[\tilde{R}\omega_{m}^{\times}]-\text{tr}[\tilde{R}\beta^{\times}]+\text{tr}[\tilde{R}\hat{\beta}^{\times}]\nonumber\\
 &\quad+\text{tr}[\tilde{R}(B_1\delta_{1})^{\times}]-\text{tr}[\tilde{R}(G_{q}\psi(\tilde{R}\epsilon))^{\times}]\nonumber\\
 &=-\text{tr}[\tilde{R}\tilde{\beta}^{\times}]+\text{tr}[\tilde{R}(B_1\delta_{1})^{\times}]\nonumber\\
 &\quad-\text{tr}[\tilde{R}(G_{q}\psi(\tilde{R}\epsilon))^{\times}]\nonumber\\
\implies\text{tr}[\dot{\tilde{R}}] &=(\tilde{R}-\tilde{R}^{\top})^{\vee}\cdot(\tilde{\beta}-B_1\delta_{1}+G_{q}\psi(\tilde{R}\epsilon)).\label{eq:trace_Rtilde_dot}
\end{align}
Using \eqref{eq:trace_Rtilde_dot}, we express $\dot{e}_R$ as:
\begin{align*}
\dot{e}_{R}&=e_R^{\times}(\omega_{m}-\hat{\beta})+\frac{2}{\sqrt{1+\text{tr}[\tilde{R}]}}E_c(B_1\delta_{1}-\tilde{\beta})\\
&\quad-\frac{2}{\sqrt{1+\text{tr}[\tilde{R}]}}E_c^{\top}G_{q}\psi(\tilde{R}\epsilon)\\
&\quad-\frac{[e_{R}\cdot\tilde{\beta}-e_{R}\cdot(B_1\delta_{1})+e_{R}\cdot G_{q}\psi(\tilde{R}\epsilon)]}{2\sqrt{1+\text{tr}[\tilde{R}]}}e_{R}
\end{align*}
Re-arranging terms, it follows that:
\begin{align*}
\dot{e}_{R}&=e_R^{\times}(\omega_{m}-\hat{\beta})-\frac{2}{\sqrt{1+\text{tr}[\tilde{R}]}}\left(E_c^{\top}+\frac{1}{4}e_{R}e_{R}^{\top}\right)G_{q}\psi(\tilde{R}\epsilon)\\
&\quad+\frac{2}{\sqrt{1+\text{tr}[\tilde{R}]}}\left(E_c+\frac{1}{4}e_Re_R^{\top}\right)(B_1\delta_{1}-\tilde{\beta})
\end{align*}
Finally, substituting $E(\tilde{R})$ from \eqref{eq:E_Rtilde}, we obtain \eqref{eq:eR_dot}.
\end{proof}

\bibliographystyle{IEEEtran}
\bibliography{Library}

\end{document}